\documentclass{aa}  
\usepackage{graphicx}
\usepackage{txfonts}
\usepackage{fixltx2e}
\newcommand{\xmm}{{XMM-{\em Newton} }}
\newcommand{\xmmns}{{XMM-{\em Newton}}}
\newcommand{\swift}{{\em Swift }}
\newcommand{\swiftns}{{\em Swift}}
\newcommand{\srcnamelong}{{XMMSL1~J061927.1-655311 }}

\newcommand{\srcname}{{XMMSL1~J0619-65 }}
\newcommand{\srcnamens}{{XMMSL1~J0619-65}}
\newcommand{\srcnamemlong}{{2MASX~06192755-6553079 }}
\newcommand{\srcnamemlongns}{{2MASX~06192755-6553079}}

\newcommand{\sdssnine}{{SDSS~J095209.56+214313.3 }}
\newcommand{\swtd}{{SWIFT~J164449.3+573451 }}
\newcommand{\fluxUnits}{{ergs s$^{-1}$cm$^{-2}$ }}
\newcommand{\fluxUnitsns}{{ergs s$^{-1}$cm$^{-2}$}}
\newcommand{\lumUnits}{{ergs s$^{-1}$ }}
\newcommand{\lumUnitsns}{{ergs s$^{-1}$}}

\newcommand{\cred}{{$C_{r}$ }}
\begin{document}
   \title{An X-ray and UV flare from the galaxy \srcnamelong}

   \subtitle{}

   \author{R.D. Saxton
          \inst{1}
          \and
          A.M. Read\inst{2}
          \and
          S. Komossa\inst{3}
          \and
          P. Rodriguez-Pascual\inst{1}
          \and
          G. Miniutti\inst{4}
          \and
          P. Dobbie\inst{5}
          \and
          P. Esquej\inst{4,6}
          \and
          M. Colless\inst{7}
          \and
          K. W. Bannister\inst{8,9}
          }

   \offprints{R. Saxton}

   \institute{XMM SOC, ESAC, Apartado 78, 28691 Villanueva de la Ca\~{n}ada, Madrid
              , Spain\\
              \email{richard.saxton@sciops.esa.int}
         \and
             Dept. of Physics and Astronomy, University of Leicester, Leicester LE1 7RH, U.K.
         \and
             Max Planck Institut f\"ur Radioastronomie, Auf dem Huegel 69, 53121 Bonn, Germany
         \and
             Centro de Astrobiolog\'ia Depto. Astrofisica (INTA-CSIC), ESAC campus, Apartado 78, 28691 Villanueva de la Ca\~{n}ada, Spain
         \and
             School of Physical Sciences, University of Tasmania, Hobart, TAS, 7001, Australia
         \and
             Departamento de Astrofísica, Facultad de CC. Físicas, Universidad Complutense de Madrid, E-28040, Madrid, Spain
	 \and
	     Research School of Astronomy and Astrophysics, Australian National University, Canberra, ACT 2611, Australia
         \and
             CSIRO Astronomy and Space Science, PO Box 76, Epping NSW 1710, Australia
         \and
             Bolton Fellow
        }

   \date{Received September 15, 1996; accepted March 16, 1997}

  \abstract
   {}
   {New high variabililty extragalactic sources may be identified by comparing
the flux of sources seen in the \xmm Slew Survey with detections and upper
limits from the ROSAT All Sky Survey.}
   {Detected flaring extragalactic sources, are subsequently monitored 
with \swift and \xmm to track their temporal and spectral evolution.
Optical and radio observations are made to help classify the galaxy,
investigate the reaction of circumnuclear material to the X-ray flare
and check for the presence of a jet.}
   {In November 2012, X-ray emission was detected
from the galaxy \srcnamelong (a.k.a. \srcnamemlongns), a factor 140 times higher than an upper limit
from 20 years earlier. Both the X-ray and UV flux subsequently fell, over
the following year, by factors of 20 and 4 respectively. Optically, the
galaxy appears to be a Seyfert I with broad Balmer lines and weak, narrow,
low-ionisation emission lines, at a redshift of 0.0729. The X-ray luminosity
peaks at L$_{X}\sim8\times10^{43}$ \lumUnits
with a typical Sy I-like power-law X-ray spectrum of $\Gamma\sim2$.
The flare has either been caused by a tidal disruption event
or by an increase in the accretion rate of a persistent AGN.}
   {}


   \maketitle
%
%________________________________________________________________

\section{Introduction}

Galaxy nuclei showing large X-ray variability, in excess of a factor
100, are rare objects. In comparisons of the ROSAT (\cite{Trumper}) All
Sky Survey (RASS; Voges 1999) and ROSAT pointed data, 6 such events
were found out of the many thousands of galaxies observed.
Of these, four (NGC 5905: \cite{Bade96}; RXJ~1242.6-1119: \cite{Komossa99b}; 
RXJ~1420.4+5534: \cite{Greiner} and RXJ~1624.9+7554: \cite{Grupe99}) 
were interpreted as tidal
disruption events (\cite{Hills}; \cite{Luminet}; \cite{Rees88}) having essentially no 
evidence for permanent AGN activity and fading in X-rays until they became
undetectable. The other two (WPVS~007: \cite{Grupe95}; IC 3599: \cite{Grupe95b}; 
\cite{Brandt}; \cite{Komossa99}) showed optical emission-line spectra of classical
AGN and were interpreted as changes in emission or absorption from a persistent AGN.

The XMM-Newton Slew Survey (XSS; \cite{Saxton08}) represents a new opportunity
to investigate high-amplitude flux changes in extragalactic sources over timescales
of years. Three tidal disruption candidates 
(NGC~3599 and SDSS~J1323+48: \cite{Esquej07}; \cite{Esquej08}; SDSS~J1201+30: \cite{Saxton12}) and one highly 
variable persistent AGN (GSN~069: \cite{Miniutti13}) have 
previously been reported.

In November 2012, \xmm discovered strong X-ray emission coming from the
vicinity of \srcnamemlongns; a galaxy which has not been detected
previously in the RASS or other X-ray surveys. An upper limits analysis of
the RASS data shows that the soft X-ray flux was at least 140 
times higher
in the \xmm observation than it was 20 years earlier. Here we present an
analysis of follow-up observations of this source.

The paper is structured as follows:
in Section 2 we discuss the flare detection and source identification;
in Sections 3, 4 \& 5 we present X-ray, UV, optical and radio follow-up 
observations; in Sections 6 \& 7 we perform a temporal and spectral analysis
of the source and in Section 8 we assess whether the flare
characteristics can be explained by an AGN or a TDE.
The paper is summarised in Section 9.

A $\lambda$CDM cosmology with ($\Omega_{M},\Omega_{\Lambda}$) = (0.27,0.73)
and  $H_{0}$=70 km$^{-1}$s$^{-1}$ Mpc$^{-1}$ has been assumed throughout.

\section{X-ray flare identification}
During the slew 9236700005, performed on November 12$^{th}$ 2012, \xmm 
(\cite{jansen}) detected
a source, \srcnamelong (hereafter \srcnamens), with an EPIC-pn, medium filter,
 0.2--2 keV count rate of 
$2.1\pm0.5$ count s$^{-1}$. This corresponds to a flux of 
$F_{0.2-2}\sim3.0\times10^{-12}$ \fluxUnits using a 
simple spectrum of a power-law
of index 2, absorbed by the Galactic column (see
sections below) and a full-band flux $F_{0.2-10}\sim5\times10^{-12}$ \fluxUnitsns.  
We calculate a 2-sigma upper limit from the RASS 
 at this position of
0.0019 count s$^{-1}$ (see \cite{Esquej07} for a description of the upper limit
calculation); a factor 140 lower flux using the same spectral model. 

The source position lies $4.2\arcsec$ from the galaxy 
\srcnamemlong (\cite{Skrut}) which with 
J=$14.43\pm{0.10}$, H=$13.92\pm{0.14}$, K=$13.62\pm{0.20}$, R=14.4 is the 
only bright optical and infrared source within the $8\arcsec$ error circle (\cite{Saxton08}).

%In archival data it was found that this position had been observed in four
% earlier slews yielding 0.2--2 keV count rates of 
%$<0.37$ count s$^{-1}$ in 2007-07-07, $<0.75$ count s$^{-1}$ in 2011-10-13,
%$1.0\pm0.2$ count s$^{-1}$ from 2012-06-01 and $<0.67$ count s$^{-1}$ in 
%2012-06-22.

A crude analysis may be performed on the 21 photons in the 
slew spectrum to investigate the gross spectral properties of the 
detection. Detector matrices are calculated, taking into account 
the transit of the
source across the detector, using a technique
outlined in \cite{Read08}. The source has a power-law 
slope $\sim 2$, assuming no intrinsic
absorption above the Galactic value of $4.4\times10^{20}$cm$^{-2}$ (\cite{Kaberla}).

\section{X-ray and UV observations}
\begin{figure}
\centering
\rotatebox{0}{\includegraphics[height=6.5cm]{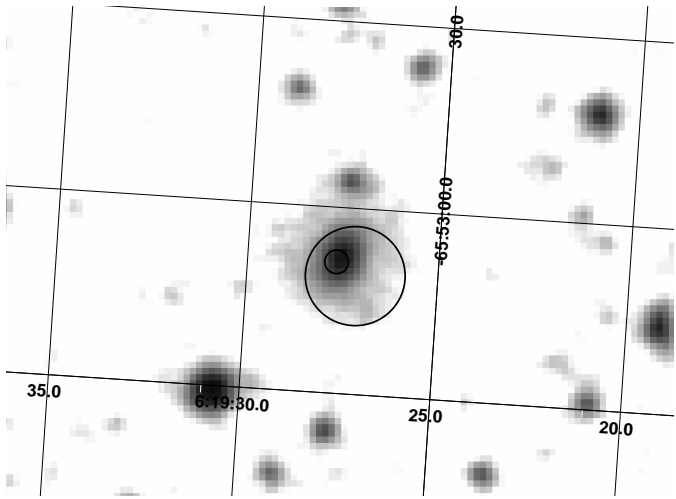}}
\caption[dss image]
{ \label{fig:swiftimage} An r-filter image of the galaxy from the digitised
sky survey, 
 shown with the \xmm slew error circle ($8\arcsec$ radius) and UVOT-enhanced
\swift error circle ($1.9\arcsec$ radius; see text) centred on the detections.
}
\end{figure}

An X-ray monitoring program was initiated with \swift to 
follow the evolution of the source flux and spectrum. Snapshot 3ks observations
were made, initially once a week and then less frequently, with the \swiftns-XRT 
(\cite{Burrows05}) in photon
counting mode and the UV optical telescope (UVOT; \cite{Roming}),
using the filter of the day. The \swiftns-XRT observations have been analysed following the 
procedure outlined in \cite{Evans} and the UVOT data have been 
reduced as described in \cite{Poole}. 
An accurate position for the source in the \swiftns-XRT field can be determined
by matching the UVOT field of view with the USNO-B1 catalogue and registering
the XRT field accordingly (\cite{Goad}). The resulting source position is
then constrained to a 1-sigma radius of $1.9\arcsec$ and is coincident with
the galactic nucleus (see Fig.~\ref{fig:swiftimage}).

In parallel, two 30ks \xmm pointed observations were triggered
on 2012-12-15 (obsid=0691100201) and 2013-04-20 (obsid=0691100301).
In each observation, the EPIC-pn and MOS-1 cameras 
were operated in full frame mode with the thin1 filter in place,
while the MOS-2 camera was used in small window mode with the medium 
filter. The source was too
faint for statistically significant data to be collected from the
reflection grating spectrometers. 
 
The XMM data were analysed with the \xmm Science Analysis System 
(SAS v13.0.0; \cite{Gabriel}). Light curves were extracted from the 
observations and searched for periods of high background flaring.
The first observation was clean but the last 8ks of the second observation 
were affected by strong background flares and were excluded from the analysis.
A list of all X-ray observations and useful exposure times is given in Table.~\ref{tab:xobs}.

%
% Following fluxes have been calculated from XSPEC using a
% model of NH=4.4E20, slope=2. Swift data fitted in range 0.3-2 keV,
% XMM in range 0.2-2 keV.
% Count rates have been removed from the original table which can 
% be found at the end of this file
%
\begin{center}
%\begin{table*}[ht]
\begin{table}
{\small
\caption{X-ray observation log of \srcname}
\label{tab:xobs}      % is used to refer this table in the text
\hfill{}
\begin{tabular}{l c l l}
\hline\hline                 % inserts double horizontal lines
Mission$^{a}$ & Date & Exp time$^{b}$ & Flux$^{c}$ \\
              &      &   (s)    &    ($10^{-12}$\fluxUnitsns) \\
\\
ROSAT     & 1990          & 4470  &  $<0.021$  \\
XMM slew  &   2007-07-07  & 8.5  & $<0.52$   \\
XMM slew  &   2011-10-13  & 4.0  & $<1.05$   \\
XMM slew  &   2012-06-01  & 9.3  & $1.5\pm{0.6}$   \\
XMM slew  &   2012-06-16  & 7.8  & $<1.44$   \\
XMM slew  &   2012-06-22  & 4.5  & $<0.94$   \\
XMM slew  &   2012-11-12  & 9.9  & $3.0\pm{1.0}$   \\
XMM slew  &   2012-11-14  & 3.6  & $2.9\pm{0.9}$   \\
XMM slew  &   2012-11-21  & 3.9  & $2.4\pm{0.9}$ \\
SWIFT  & 2012-11-25       &  3230   & $2.09\pm{0.21}$ \\
XMM pointed & 2012-12-15  &  26600  & $1.18\pm{0.15}$ \\
SWIFT  & 2012-12-27      &  3450   & $0.70\pm{0.12}$ \\
SWIFT  & 2013-01-02       &  2960  & $0.57\pm{0.12}$ \\
SWIFT  & 2013-01-09       &  2870  & $0.63\pm{0.15}$ \\
SWIFT  & 2013-01-16       &  2890  & $0.38\pm{0.12}$ \\
SWIFT  & 2013-01-23       &  2510  & $0.24\pm{0.12}$ \\
SWIFT  & 2013-02-06       &  2550  & $0.50\pm{0.12}$ \\
SWIFT  & 2013-03-08       &  1920  & $0.26\pm{0.11}$ \\
SWIFT  & 2013-04-11       &  1940  & $0.56\pm{0.14}$ \\
XMM pointed & 2013-04-20  &  21600 & $0.41\pm{0.01}$\\
XMM slew & 2013-04-25     &  10.2  & $0.6\pm{0.3}$\\
SWIFT  & 2013-05-08       &  1890  & $0.23\pm{0.11}$ \\
SWIFT  & 2013-06-08       &  1980  & $0.75\pm{0.20}$ \\
XMM slew & 2013-06-20     &  8.4   & $<1.08$ \\
SWIFT  & 2013-07-08       &  1920  & $0.25\pm{0.11}$ \\
SWIFT  & 2013-12-04       &  4420  & $0.12\pm{0.05}$ \\
\hline                        % inserts single horizontal line
\end{tabular}}
\hfill{}
\\
\\
$^{a}$ \xmmns, EPIC-pn camera: slew observations performed in {\em full frame}
mode with
the Medium filter; pointed observations performed in {\em full frame} mode with
the thin1 filter. \swiftns-XRT observations performed in {\em pc} mode. \\
$^{b}$ Useful exposure time after removing times of high background flares. \\
$^{c}$ Absorbed flux in the 0.2-2 keV band. For simplicity this has been calculated 
using a power-law model of slope 2.0 and 
Galactic absorption of $4.4\times10^{20}$cm$^{-2}$ in all cases.\\
%\end{table*}
\end{table}
\end{center}

\section{Optical observations}
We observed \srcname with the AAOmega 2dF spectrograph
(\cite{Saunders04}; \cite{Sharp06}) at the prime focus of the Anglo–
Australian Telescope on 2012-12-03 for 3 x 600 s, in order to obtain an optical
classification of the source and to search for transient emission lines excited by the flare. We used the 580V (3700--5800 °A) and 385R (5600--8800 °A) low-resolution
gratings (R  1300). The 2dF data were reduced using the 2dFDR
software of the AAO (e.g. \cite{SharpBirch}) which performs
bias subtraction, fibre–flat fielding and wavelength calibration in an
automated manner. The response curves for the blue and red arms
were determined with an observation of several standards 
and were then used to obtain a relative flux calibration of the
spectrum of \srcnamens. The blue spectrum was then renormalised to the
B band (4500 \AA) flux from the first \xmm pointed observation of 
2012-12-15 and the red spectrum scaled to match the blue in the 
overlapping area.

\begin{figure*}[t]
\centering
%\rotatebox{0}{\includegraphics[height=7cm]{2MASX_619_optical.ps}}
%\rotatebox{-90}{\includegraphics[height=9cm]{optical_doubleplot_v2.ps}}
%\rotatebox{-90}{\includegraphics[height=9cm]{optical_spectrum.ps}}
\rotatebox{-90}{\includegraphics[height=12cm]{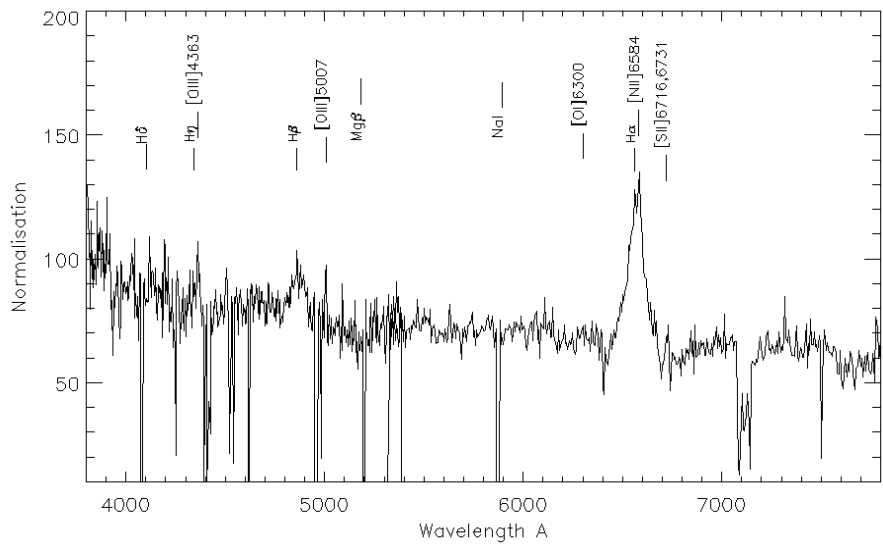}}
\caption[AAT spectrum]
{ \label{fig:AATspec}AAOmega 2dF spectrum of \srcname
taken on 2012-12-03, after removal of narrow skyline features.} 

\end{figure*}

The spectrum shows broad H$\alpha$ and H$\beta$ and weak, narrow,
low-ionisation Balmer lines and lines of [OIII], [NII] and [SII] (Fig.~~\ref{fig:AATspec}). 
In table~\ref{tab:optlines} we give the fluxes and relative intensities 
of detected emission lines.  
The diagnostic narrow line ratios [OIII]$\lambda$5007/H$\beta$=$3.1\pm{1.3}$; 
[NII]$\lambda$6583/H$\alpha$=$0.85\pm{0.15}$
 and [SII]$\lambda$$\lambda$6716,6713/H$\alpha$=$1.0\pm{0.3}$, classify
the source as an AGN (\cite{veillandost}) or place it, more precisely, on
the boundary between a Liner and a Seyfert galaxy (\cite{Ho08}) . The FWHM of the Balmer lines and
their strengths, relative to the narrow lines, identify this as a Seyfert I
galaxy (\cite{Seyfert43}).
The source redshift as determined from
the positions of narrow H$\alpha$ and [NII] is z=$0.0729\pm{0.0002}$.

\begin{figure}
\centering
\rotatebox{-90}{\includegraphics[height=8cm]{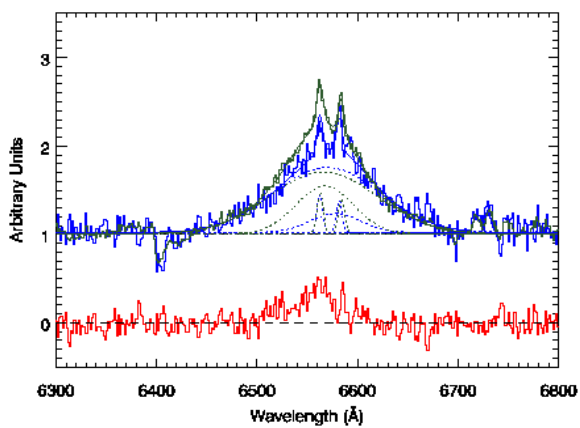}}
\caption[The line profile of $H\alpha$ in \srcnamens.]
{ \label{fig:halphaspec} A comparison of the $H_{\alpha}$ profile
of the two AAOmega 2dF spectra of \srcname
taken on 2012-12-03 (blue) and 2013-10-18 (green). Both spectra have
been normalised to their respective continua. The dotted lines represent fits
of a multi-component (very broad, broad and narrow lines) model (see text). 
The lower panel shows the
difference spectrum produced when subtracting the normalised 2012-12-03
profile from that of 2013-10-18.}

\end{figure}

A second observation was made with the same detector configuration on
2013-10-18 for 1200 s. This shows the same 
qualitative line features as 
the first observation, namely broad Balmer lines and weak narrow lines. 
While a direct comparison of line fluxes, between the two observations, is 
prohibited by the lack of an absolute flux calibration for the
second spectrum, a comparison of the broad H$_{\alpha}$ profiles can be made
by normalising both spectra to the continuum (Fig.~\ref{fig:halphaspec}).
The equivalent width of the broad H$_{\alpha}$ line is greater in the second observation. If the two profiles are subtracted, after normalisation to the continuum, then the residual profile has a width of FWHM$\sim3000$ km/s, intermediate between the very broad, FWHM$\sim6000$ km/s, component and the narrow line. 
Fitting the profiles separately, with a three-component model, gives a consistent strength for the very broad and narrow lines, while the medium width line doubles in intensity between the first and second observations.

\begin{center}
%\begin{table*}[ht]
\begin{table}
{\small
\caption{Optical line widths and intensities for \srcname from the AAOmega
2dF observation of 2012-12-03.}
\label{tab:optlines}      % is used to refer this table in the text
\hfill{}
\begin{tabular}{l c c c c}
\hline\hline                 % inserts double horizontal lines
Line & Rest $\lambda$ & Intensity relative & FWHM & Flux$^{a}$ \\
     &   \AA    &    to narrow H$\alpha$ & km/s & $10^{-16}$ ergs s$^{-1}$ cm$^2$ \\
\\
% $[OIII]$ & 3727.0 &  2.72   & 1.53 & 515.4\\
%$[$OIII$]$ & 4363.0  & 2.85  &  1.08 & 461 & 9.4 \\
 H$\beta$ & 4862  &  0.37 &  300 &  $2.8\pm{1.7}$ \\
 H$\beta$ & 4872 & 8.24  & 5174  & $66.8\pm{16.7}$ \\
 $[$OIII$]$ & 5007  & 1.14   &  330 & $8.6\pm{2.2}$ \\
%\hline
%\hline
%   & 5875  & 3.50 & 1.93  &  420.7 & 17 \\
 %$[$NII$]$  & 6548  & 3.50 & 0.40  &  377 & 3.5 \\
 H$\alpha$  & 6563  & 1.00 & 293 & $8.1\pm{1.0}$ \\
 H$\alpha$ & 6570  &  33.54 &  5716 & $271.9\pm{15.5}$ \\
 H$\alpha$ & 6578  &  4.61 &  2687 & $37.3\pm{9.7}$ \\
$[$NII$]$  & 6582  & 0.85 & 292 & $6.9\pm{0.9}$ \\
 $[$SII$]$  & 6719  & 0.42 &  286 & $3.4\pm{0.5}$ \\
 $[$SII$]$  & 6730  & 0.58  &  285 &  $4.7\pm{0.6}$\\
%   & 7319  & 3.50 & 0.91  &  338 & 8.0 \\
\hline                        % inserts single horizontal line
\end{tabular}}
\hfill{}
\\
\\
$^{a}$ The quoted errors on the flux represent the statistical error.  The estimated systematic error on the absolute line fluxes is $\sim25\%$. \\

%\end{table*}
\end{table}
\end{center}

%\subsection{Catalina sky survey}
%Monitoring data in the V band exists for this source from 2010 to May 2012 from
%the Catalina sky survey (\cite{catalina}).
%The source shows variations of $\delta m_{v} \sim 0.2$ which indicate that it is an 
%AGN. Or on the contrary that it isn't an AGN. LEARN WHAT THIS DOES MEAN.

\section{Radio observation}

We observed \srcname with the ATCA CABB system (\cite{Wilson11}) on
2013-01-21 for 5 hours staring at 07:35 UT, in order to detect possible jet
emission related to the flare or to search for low-level radio emission from 
a permanent AGN. The array was in the 750C
configuration. We observed with 2 GHz bandwidth for 5 bands centered at
2.1 GHz, 5.5 GHz, 9.0 GHz, 17 GHz and 19 GHz. The 2.1 GHz band was badly
affected by RFI, so that only approximately 2.5 GHz was usable. We
flagged and calibrated the data according to standard procedure in {\sc
Miriad} (\cite{Sault}). We used PKS1934-638 as a primary calibrator for
the lowest 3 bands, and PKS2221-052 for the upper two bands using 3.894
Jy at 17 GHz and 4.159 Jy at 19 Ghz. Regular phase calibration scans
were taken of the nearby calibrator (0515-674) for the lower 3 bands,
and 0623-6436 for the upper 2 bands. We obtained 3-4 cuts per band at
roughly equally spaced hour angles over 5 hours.

As the array was in a relatively short configuration, with a large gap
in $uv$ coverage between the first 5 antennas and antenna 6, we removed
antenna 6 during the imaging. Images were made with multifrequency
synthesis and natural weighting to maximise sensitivity to point
sources, and cleaned with multifrequency clean. For the 5.5 and 9 GHz
data, we also performed one round phase self-calibration based on point
sources that were in the field.

XMMSLJ0619-6553 was not detected in any band above 5 sigma. The 5 sigma
upper limits were 2.1 GHz: 984 $\mu$Jy; 5.5 GHz: 110 $\mu$Jy; 
9.0 GHz: 150 $\mu$Jy; 17 GHz: 340 $\mu$Jy; 19 GHz: 486 $\mu$Jy.

\section{X-ray and UV variability}
\begin{figure}
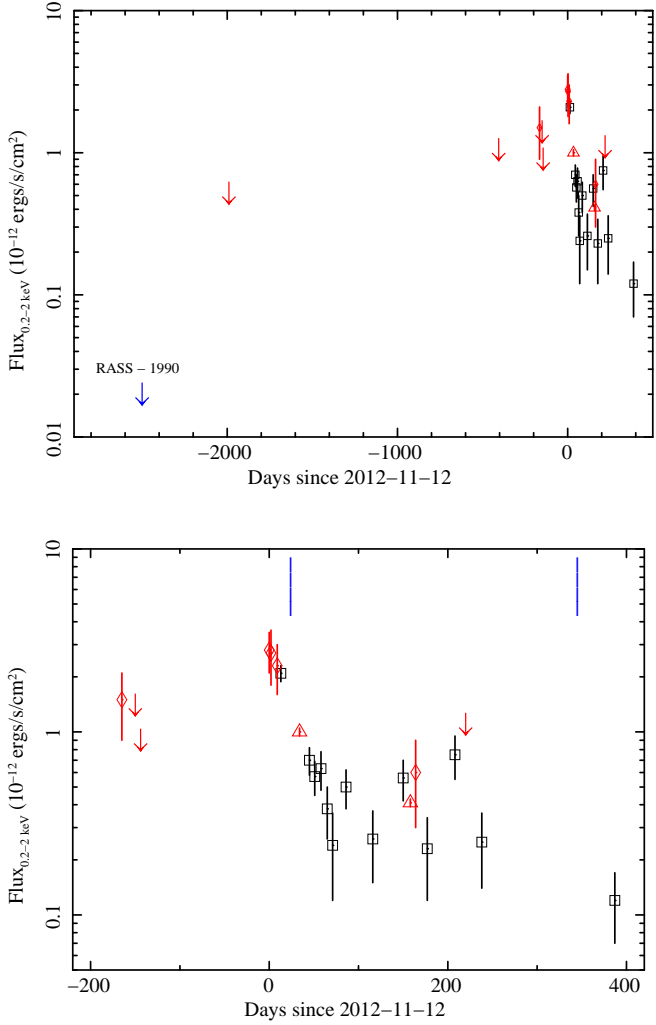

\begin{center}
\begin{minipage}{3in}
\hspace*{-0.5in}    \rotatebox{-90}{\includegraphics[height=9cm]{lcurve_long_full.ps}}
\end{minipage}

\vspace*{0.2in}

\begin{minipage}{3in}
\hspace*{-0.5in}   \rotatebox{-90}{\includegraphics[height=9cm]{lcurve_long_recent.ps}}
\end{minipage}

\vspace*{0.2in}

  \end{center}
\caption[\srcname light curve]
{ \label{fig:lcurve} The 0.2--2 keV X-ray light curve of \srcnamens.
Upper: the full historical light-curve, lower: a zoom into the
more recent data. The symbols are: XSS (red diamonds or arrows for upper limits), \xmm pointed observations (red triangles), \swiftns-XRT (black squares),
ROSAT upper limit (blue arrow). The blue dotted vertical lines indicate the 
dates of the optical observations.}
\end{figure}

In figure~\ref{fig:lcurve} we show the historical X-ray 
light curve for \srcnamens. Around 2012-11-12 the source flux apparently 
peaked, reaching a value 140 times higher than an upper limit that can be derived from 
RASS data taken in 1990. It 
then experienced a decline of a factor 20 while continuing to exhibit 
variability of a factor few between monthly observations.

\begin{figure}
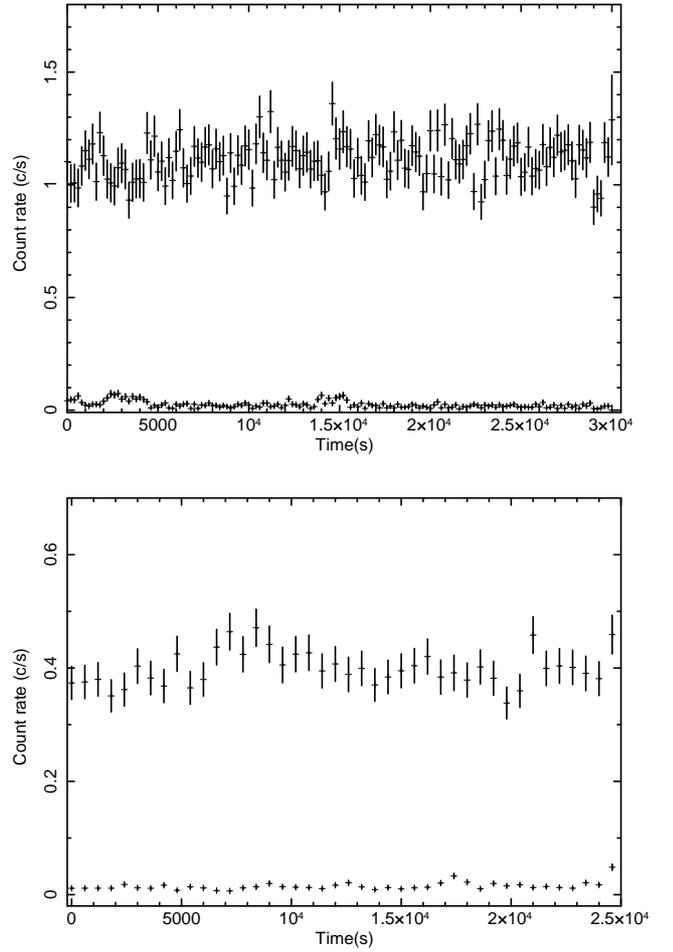

  \begin{center}
\begin{minipage}{3in}
    \rotatebox{-90}{\includegraphics[width=6cm]{short_lcurve1.ps}}
\end{minipage} 

\vspace*{0.2in}

\begin{minipage}{3in}
    \rotatebox{-90}{\includegraphics[width=6cm]{short_lcurve2.ps}}
\end{minipage} 

\vspace*{0.2in}

  \end{center}
\caption[\srcname Short term light curve]
{ \label{fig:lcurve_short} The background subtracted, exposure corrected, 
EPIC-pn, 0.2--10 keV, light curves and area-normalised background, for the 2012-12-15 (top) 
and 2013-04-24 (bottom) \xmm pointed observations.}
\end{figure}

In Fig.~\ref{fig:lcurve_short}  we show the short-term light curve from the 
two \xmm pointed observations.
The X-ray flux shows small variations, with an amplitude up to 25\%, on 
time scales of a few thousand seconds.

%Short-term light curves

\subsection{UV light curve}

During the two pointed \xmm observations, the optical monitor (OM)
cycled between the {\it B}, {\it U}, {\it UVW1} and {\it UVM2} filters.
\swiftns-UVOT observations were
performed with the filter of the day, either the {\it uv, uvw1, uvm2 or uvw2} filters,
except for the last observation of 2013-12-04 which cycled through all the filters.
The galaxy was detected in all of the filters.
Relative filter fluxes were determined using several nearby sources of
comparable brightness as references. The absolute flux scale was taken from 
the \swiftns-UVOT filters, with \xmmns-OM points scaled to these by 
filter-dependent factors of $\sim$2--3, using the technique outlined 
in \cite{Grupe08}.

\begin{figure}
\centering
%\rotatebox{-90}{\includegraphics[height=9cm]{optfilt.ps}}
\hspace*{-0.2in} \rotatebox{0}{\includegraphics[height=7cm]{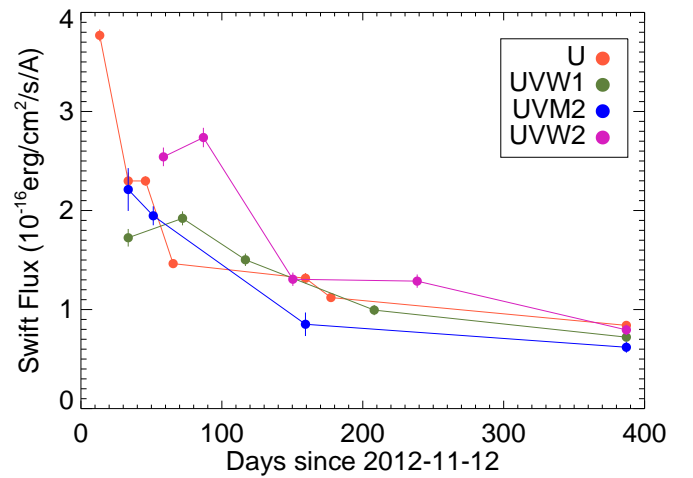}}
\caption[UV light curve of \srcnamens.]
{ \label{fig:uvlcurve} The UV light curve of \srcname
taken from the \swiftns-UVOT and \xmmns-OM observations using the filters:
{\it U} (3480\AA), {\it UVW1} (2950\AA), {\it UVW2}
(2120\AA) and {\it UVM2} (2340\AA)}
\end{figure}

In figure~\ref{fig:uvlcurve} we see that the flux, in all the UV filters, 
has responded to a strong flare coeval with the X-ray peak.
After the flare the UV flux declined in all the filters, with the better
monitored U band (3480\AA) reducing by a factor $4.2\pm{0.3}$ or 
$1.56\pm{0.08}$ magnitudes in 374 days.   

Galex observed the galaxy on 2007-03-03 finding an NUV (2267 \AA)
%mag of $20.24\pm{0.12}$ and FUV (1516 \AA) mag of $20.63\pm{0.13}$.
flux of $1.7\pm{0.2}\times10^{-16}$ ergs s$^{-1}$cm$^{-2}\AA^{-1}$ 
and FUV (1516 \AA) flux of $2.6\pm{0.3}\times10^{-16}$ 
ergs s$^{-1}$cm$^{-2}\AA^{-1}$. It reobserved the galaxy in the
NUV only on 2011-10-26 and found a consistent flux.
The NUV filter has a similar bandpass to the uvw2 filter and
its flux is close to that of the third and fouth observations made 
by \swift with that filter.

\subsection{X-ray spectral analysis}
We performed a spectral analysis of the two \xmm pointed observations,
taken 34 and 159 days after the peak flux seen on 2012-11-12 
(see Fig.~\ref{fig:lcurve}). 
We extracted spectra for the three \xmm EPIC cameras with the \xmm SAS
task {\tt especget} which uses optimum extraction regions
and a local background. Fits were performed simultaneously on the ungrouped
spectra, using the Cash statistic (\cite{Cash}), over the energy range  
0.3--10 keV, using a constant to account for
the small differences in normalisation between the instruments.
The first \xmm pointed observation (hereafter XMM1) had 31200 background 
subtracted counts in the three cameras and in this energy range, while 
the second pointed observation 
(hereafter XMM2) had 12300 background subtracted counts. Quoted errors
are 90\% confidence unless otherwise stated.
 
\begin{figure}
\centering
\rotatebox{-90}{\includegraphics[height=9cm]{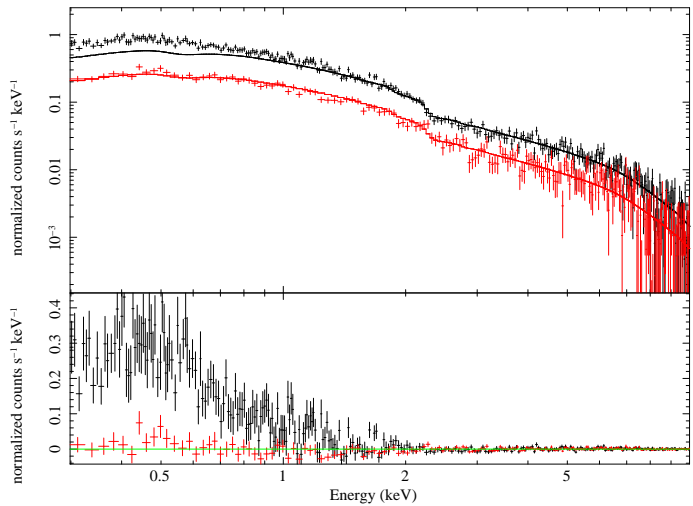}}
\caption[A comparison of spectra from the two \xmm pointed observations.]
{ \label{fig:doublespec} A comparison of EPIC-pn spectra from the \xmm pointed 
observations of 2012-12-15 (black) and 2013-04-20 (red). The lower panel
presents residuals to independent fits of a power-law model, 
absorbed by the Galactic column, to the 2-10 keV data of each spectrum.
}
\end{figure}

We applied a series of spectral models to XMM1 and XMM2 both separately
and jointly.
As a first step we fit the XMM1 observation with a simple power-law model 
and galactic absorption of $4.4\times10^{20}$cm$^{-2}$.
The fit is reasonably good (C/dof=\cred=3612/3230) and yields a slope of 
$1.96\pm{0.02}$ 
consistent with a typical Seyfert I spectrum (e.g. \cite{NandraPounds94}).
Fitting in the 0.3--2 keV and 2--10 keV ranges independently gives slopes of 
$2.1\pm{0.03}$ and $1.80\pm{0.03}$ respectively, showing that there is curvature in the spectrum (see Fig.~\ref{fig:doublespec}). 
We tried to model the curvature as a low-energy excess 
attributable to a thermal plasma or more successfully to a black-body.
The latter improved the fit (\cred=3490/3228) and gave a temperature, 
$kT=114^{+8}_{-29}eV$, consistent with the ubiquitous 100-200 eV seen when 
fitting the soft excess of a wide range of AGN (\cite{GierlinskDone}). 
Residuals remain around 6 keV which 
can be modelled by a narrow gaussian with energy fixed at 6.4 keV in the source rest frame;
further improving the fit (\cred=3480/3227) and giving a line strength 
of $1.40^{+0.72}_{-0.77}\times10^{-6}$ photons cm$^{-2}$ s$^{-1}$ and equivalent width of $79^{+65}_{-39} eV$.

XMM2 has a count rate 2.5 times lower than XMM1.
Independently fitting a power-law in the ranges 0.3--2 keV and 2--10 keV yields slopes
of $1.88\pm{0.04}$ and $1.79\pm{0.09}$ implying that the spectral shape has changed (Fig.\ref{fig:doublespec}).
The simple power-law plus black-body model is also improved for XMM2 
by the addition of a narrow 
Fe K$_{\alpha}$ line of strength $6.80^{+6.53}_{-5.64}\times10^{-7}$ photons cm$^{-2}$ s$^{-1}$ and equivalent width of $93^{+89}_{-77} eV$.

An F-test gives a probability of 99.8\% and 96.1\% that the addition of
the Fe line to the XMM1 and XMM2 observations is significant, although we
note the problem with applying the test in this context (\cite{Protassov02}).

\begin{figure}
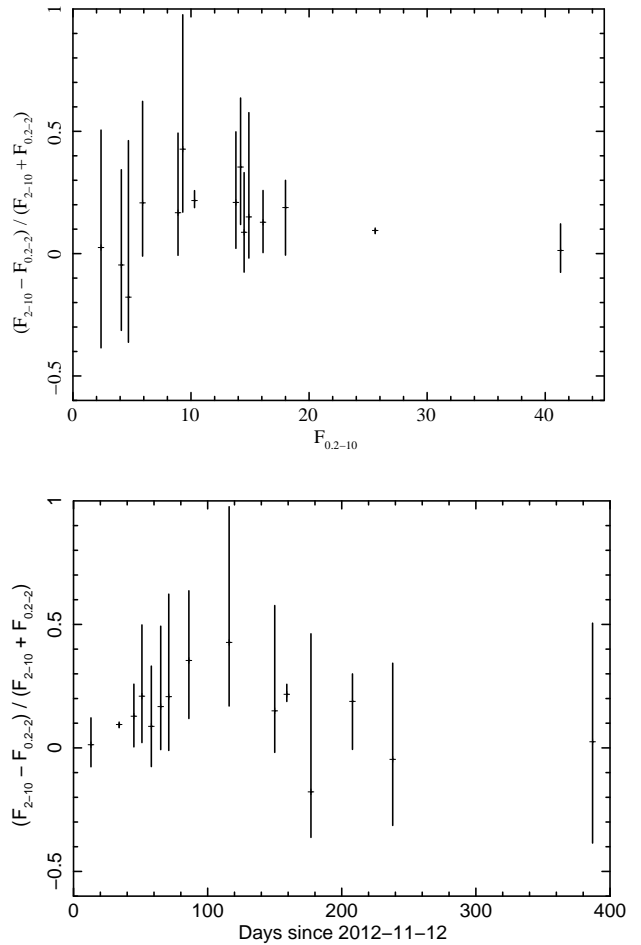

  \begin{center}
\begin{minipage}{3in}
    \rotatebox{-90}{\includegraphics[width=6cm]{hratio_flux.ps}}
\end{minipage}
 
\vspace*{0.2in}

\begin{minipage}{3in}
    \rotatebox{-90}{\includegraphics[width=6cm]{hratio_time.ps}}
\end{minipage} 

\vspace*{0.2in}

  \end{center}
\caption[\srcname hardness ratio]
{ \label{fig:hratio} The 2--10 keV to 0.2--2.0 keV flux ratio, calculated by fitting a 
phenomenological model of a power-law plus a black-body of kT=120 eV absorbed by
the Galactic column, as
a function of the (absorbed) X-ray (0.2--10 keV) flux (upper) and
observation date (lower).}
\end{figure}

To quantify how the spectral shape changes between observations we apply an absorbed power-law with a black-body of temperature kT=120 eV 
to all of our
X-ray spectra, including the \swift observations. This gives the 
relationship between flux and spectral hardness shown in Fig.~\ref{fig:hratio}. There is no obvious correlation
with flux, within the errors. Possibly the source has hardened with time and
then subsequently softened.

To consistently describe the neutral iron emission line and the
spectral curvature we added a model for reflection from distant material.
We adopted the {\tt pexmon} model (\cite{Nandra07}), which considers
Compton reflection and the Fe K$\alpha$, Fe K$\beta$ and Ni K$\alpha$
flourescence lines, fixing abundances to the solar values. The inclination
angle was essentially unconstrained in fits and was fixed to $45\,^{\circ}$. 
Fig.~\ref{fig:doublespec} suggests that, while the continuum above 2 keV varies mainly in normalisation between the two observations, there has been a large change in the shape of the soft spectrum below 2 keV.
This can be modelled as either an increase in absorption or as a decrease in
a discrete soft emission component between the XMM1 and XMM2 observations.
In table~\ref{tab:specfits} we test these possibilities with fits to the individual and combined observations. For absorption we use a partially covering, ionized absorber (zxipcf in XSPEC; \cite{Reeves08}). The alternative, additional soft emission component, was modelled by reflection from an ionized disk ({\tt reflionx} in 
XSPEC; \cite{RossFabian}) which is relativistically blurred ({\tt kdblur} in 
XSPEC; \cite{Laor91}, \cite{Fabian2002}). We fixed the relative Fe abundance to the solar value and the power-law slope to the index of the primary continuum.
The fit statistic was found to have little dependence on the parameters of the blurring model and these were fixed to, emissivity index=3; inner radius = 4.5 $GM/c^{2}$; outer radius=100 $GM/c^{2}$ and inclination=$45\,^{\circ}$.   
In fits to the individual observations, the two models fit XMM1 equally well while the absorption model gives a somewhat better fit to XMM2 (\cred=3040/3226 compared with \cred=3055/3227).

In order to better constrain the changes between XMM1 and XMM2
we fit the two spectra simultaneously. First we looked if a simple variation of the intensity of the primary power-law, with fixed reflection and absorption could explain the data. This gave a fit statistic of \cred=6622/6457.
Then we forced the primary power-law
and distant reflection components to be constant between the two observations and allowed the
absorption alone to vary, which improved the fit to \cred=6575/6455. 
Repeating this fit but also allowing the power-law normalisation to vary reduced the statistic to \cred=6528/6454.
Finally we fixed the power-law slope and distant reflection components and 
allow the power-law normalisation and the reflection
from the disk to vary, yielding a fit parameter of \cred=6557/6455.
In this case the best fit yields a reduction in the power-law normalisation and 
an even greater reduction in the reflection normalisation. We note that this disagrees with the light-bending model (\cite{MinFab}) which argues that a reduction in power-law emission occurs due to an increase in rays which are intercepted by the disk; a mechanism which naturally leads to an increase in disk emission.
%In summary, from these spectral fits alone, we are unable to discriminate betwen variability due to increased absorption or due to decreased emission.

%\section{Interpretation as variable absorption}
%Under the assumption that flux variations between the X-ray observations are purely due
%to a partial coverer then we find an increase in NH of $5\times10^{24}$ between the two
%\xmm pointed observations. If this is extended to the RASS observation of 1990 then the covering
%factor would have had to be 100\% in this epoch to produce the observed flux reduction 
%of two orders of magnitude. The absorber(s) would also need to have covered the UV 
%emitting region.
%This should be contrasted with the negligible variations which are 
%seen in the UV magnitudes between the two \xmm pointed observations. In addition we find that the
%historical blue magnitude from a SERC plate of 1983 is consistent within the errors with the blue
%magnitude recorded by \xmm. These imply that the putative clouds covering the X-ray source are much 
%smaller than the UV emitting region. While the X-ray region is $\sim 10 r_{g}$ (e.g. ref) the UV
%region is more like xxx $r_{g}$ or $10^{xxx}$ cm. Studies of obscuring clouds in NGC 1365 \cite{risaliti}
%impied a comet-like structure of size $10^{xxx}$ cm for the nucleus and $10^{xxx}$ cm for the tail.
%It is quite plausible then that such clouds could obscure the X-rays while having little effect on 
%the UV emission.  

\renewcommand{\arraystretch}{1.5}
\begin{table*}[ht]
{\small
\caption{Spectral fits to \xmm pointed observations of \srcnamens.}
\label{tab:specfits}      % is used to refer this table in the text
\begin{center}
\begin{small}
\begin{tabular}{c c| c| c c c| c c |c}
\hline\hline                 % inserts double horizontal lines
% \multicolumn{7}{c}{emission model} & \multicolumn{3}{c}{intrinsic absorption} & C/dof  \\
\\
\multicolumn{2}{c}{Power-law$^{a}$}  & Distant Reflection$^{b}$ & \multicolumn{3}{c}{zxipcf~$^{c}$} & \multicolumn{2}{c}{Ionized Reflection$^{d}$} & C/dof~$^{e}$ \\
$\Gamma$ & Norm  &  Norm & $N_{H}$ & $xi$ & cf & $xi$ & Norm & \\
         & keV$^{-1}$ cm$^{-2}$ s$^{-1}$ & keV$^{-1}$ cm$^{-2}$ s$^{-1}$ & 10$^{22}$ cm$^{-2}$ & log & \% & log & cm$^{-2}$ s$^{-1}$ & \\
\hline\noalign{\smallskip}
\multicolumn{9}{c}{\xmm observation 1 - 2012-12-15} \\
\hline
% Power-law fit
$1.96\pm{0.02}$ & $4.84^{+0.06}_{-0.04}\times10^{-4}$ &       &     &    &    &   &  &  3612/3230  \\
% Power-law + pexmon fit
$1.98\pm{0.02}$ & $4.83^{+0.05}_{-0.04}\times10^{-4}$ & $6.14^{+2.03}_{-4.53}\times10^{-4}$ &     &     &    &   &   & 3573/3229  \\
% Power-law + pexmon + zxipcf
$2.05^{+0.03}_{-0.02}$ & $7.28^{+0.81}_{-0.84}\times10^{-4}$  & $3.86^{+1.98}_{-2.79}\times10^{-4}$ & $16.8^{+5.0}_{-6.6}$  & $1.67^{+0.77}_{-0.72}$    &  $36\pm{6}$ &    &   & 3472/3226 \\
% Power-law + pexmon + reflionx*kdblur
$1.92\pm{0.02}$   & $4.46^{+0.07}_{-0.08}\times10^{-4}$  & $2.48^{+1.66}_{-1.55}\times10^{-4}$  &    &   &   &   $10.0^{+2.7}_{-10.0}$ &  $1.03^{+0.18}_{-0.27}\times10^{-6}$ & 3471/3227  \\
\noalign{\smallskip}\hline
\multicolumn{9}{c}{\xmm observation 2 - 2013-04-24} \\
\hline\noalign{\smallskip}
% Power-law fit
$1.81\pm{0.03}$   &   $1.75^{+0.04}_{-0.03}\times10^{-4}$ &       &     &     &   &   &  &  3072/3230  \\
% Power-law + pexmon fit
$1.82\pm{0.03}$ & $1.75^{+0.04}_{-0.03}\times10^{-4}$ & $1.46{+1.36}_{-0.90}\times10^{-4}$ &     &     &    &   &   & 3064/3229  \\
% Power-law + pexmon + zxipcf
$1.76\pm{+0.04}$ & $1.93^{+0.06}_{-0.09}\times10^{-4}$  & $1.17^{+1.06}_{-0.97}\times10^{-4}$ & $500^{+0}_{-100}$  & $4.27^{+0.07}_{-0.05}$    &  $100^{+0}_{-16}$ &    &   & 3040/3226 \\
% Power-law + pexmon + reflionx
$1.79\pm{0.03}$   & $1.68^{+0.05}_{-0.06}\times10^{-4}$  & $1.13^{+1.08}_{-0.97}\times10^{-4}$  &   &  &     &   $10.0^{+31.0}_{-10.0}$ &  $2.49^{+1.54}_{-2.00}\times10^{-7}$ & 3055/3227  \\
\noalign{\smallskip}\hline
\multicolumn{9}{c}{Joint fits to the \xmm observations of 2012-12-15 and 2013-04-24$^{f}$} \\
\hline\noalign{\smallskip}
% Power-law + pexmon + zxipcf - variable plaw
$1.91^{+0.02}_{-0.02}$ & & $2.70^{+1.10}_{-1.00}\times10^{-4}$ & $65^{+64}_{-21}$  & $3.05^{+0.12}_{-0.61}$    &  $27^{+14}_{-12}$ &    &   & \\
% Norm obs1
 & $5.57^{+0.16}_{-0.10}\times10^{-4}$  &  &  &  &  &    &   &  \\
% Norm obs2
 & $2.03^{+0.64}_{-0.06}\times10^{-4}$  &  &  &  &  &    &   & 6621/6457 \\
%
% Power-law + pexmon + zxipcf
$1.90^{+0.04}_{-0.02}$ & $5.79^{+0.24}_{-0.22}\times10^{-4}$  & $2.75^{+1.90}_{-1.40}\times10^{-4}$ &   &     &   &    &   &  \\
% Power-law + pexmon + zxipcf - obs1
 &   &  & $66^{+9}_{-26}$  & $3.2^{+0.2}_{-0.4}$    &  $43\pm{11}$ &    &   & \\
% Power-law + pexmon + zxipcf -obs2
 &   &  & $180^{+80}_{-28}$  & $2.72^{+0.02}_{-0.17}$    &  $70\pm{2}$ &    &  & 6575/6455 \\
%
% Power-law + pexmon + zxipcf - variable plaw norm - TO BE RAN
$1.96^{+0.02}_{-0.02}$ &  & $2.38^{+1.30}_{-1.21}\times10^{-4}$ &   &     &   &    &   &  \\
% Power-law + pexmon + zxipcf - obs1
 &  $6.81^{+1.21}_{-1.29}\times10^{-4}$  &  & $50^{+56}_{-22}$  & $2.2^{+1.0}_{-0.2}$    &  $34^{+5}_{-15}$ &    &   & \\
% Power-law + pexmon + zxipcf -obs2
 & $2.38^{+0.25}_{-0.19}\times10^{-4}$   &  & $2.1^{+2.6}_{-1.3}$  & $-0.6^{+1.5}_{-0}$    &  $30^{+8}_{-6}$ &    &  & 6528/6454 \\
%
% Power-law + pexmon + reflionx*kdblur 
$1.87\pm{0.03}$   &  & $2.25^{+0.11}_{-0.09}\times10^{-4}$  &   &  &     &    &  & \\
% Power-law + pexmon + reflionx - obs1
 & $4.21^{+0.20}_{-0.11}\times10^{-4}$  &  &   &  &     &   $23^{+7}_{13}$ &  $3.84^{+6.20}_{-1.06}\times10^{-7}$  \\
% Power-law + pexmon + reflionx - obs2
&  $1.78^{+0.04}_{-0.05}\times10^{-4}$   &  &   &  &     &   $10^{g}$ &  $0.0^{+6.2}_{-0.0}\times10^{-8}$ & 6557/6455 \\
\noalign{\smallskip}\hline
\end{tabular}
\\
\end{small}
\end{center}
All fits include absorption by the Galactic column (model {\tt TBABS}, $N_{H}=4.4\times10^{20}$cm$^{-2}$). Errors are 90\% confidence. \\
$^{a}$ Continuum power-law with photon index and normalisation, $^{b}$ Reflection from distant matter with normalisation at 1 keV ({\tt pexmon}; \cite{Nandra07} with inclination fixed at $45\,^{\circ}{\rm C}$) , $^{c}$ ionized absorption with equivalent column density, ionization state and covering fraction, $^{d}$ ionized reflection ({\tt reflionx}; \cite{RossFabian}) with ionization state and normalisation, convolved with a relativistic blurring
model ({\tt kdblur}; \cite{Laor91}, \cite{Fabian2002}), with fixed emissivity index=3; inner radius=4.5 $GM/c^{2}$; outer radius=100 $GM/c^{2}$ and inclination=$45\,^{\circ}$,   $^{e}$ C-statistic / number of degrees of freedom. $^{f}$ Simultaneous fits to both \xmm observations; for these, the first line of each model gives the tied parameters, the second line gives the free values for the 2012-12-15 observation and the third line the free values for the 2013-04-24 observation. $^{g}$ The parameter is unconstrained.
}
\end{table*}
\renewcommand{\arraystretch}{1.0}

\section{Broadband SED}
\begin{figure}
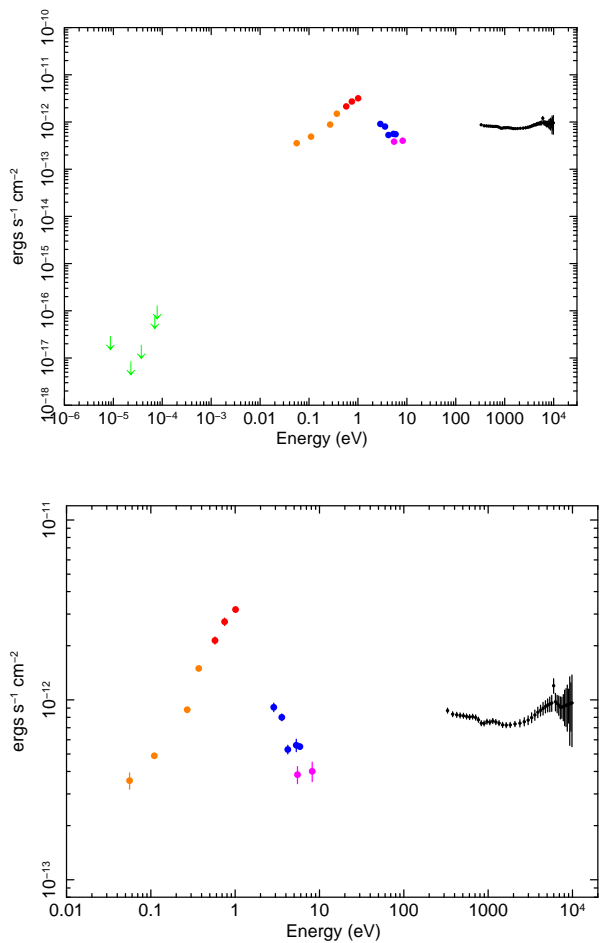

  \begin{center}
\begin{minipage}{3in}
    \rotatebox{-90}{\includegraphics[width=6cm]{sed_0619.ps}}
\end{minipage}

\vspace*{0.2in}

\begin{minipage}{3in}
    \rotatebox{-90}{\includegraphics[width=6cm]{sedzoom.ps}}
\end{minipage}

\vspace*{0.2in}

  \end{center}
\caption[Spectral energy distribution of \srcnamens.]
{\label{fig:sed} The Galactic-absorption corrected spectral energy distribution of \srcnamens. Data are from ATCA (green), WISE (orange), 2MASS (red),
\xmmns-OM (blue), Galex (pink) and \xmm EPIC-pn (black). The Galex points were
taken in 2007, likely during a slightly lower luminosity state than the \xmm
measurements of 2012-12-15. The lower panel shows a zoom into the IR-X-ray data.}
\end{figure}

In figure~\ref{fig:sed} we plot the spectral energy distribution (SED) of the
source. A measure of the relative X-ray strength can be found from the 
optical to X-ray slope, defined in \cite{just07} as 

\begin{equation}
\alpha_{OX}=0.3838 \cdot \log \left( \frac{F_{2keV}}{F_{2500\AA}}\right),
\end{equation}
where $F_{2keV}$ and $F_{2500\AA}$ are the monochromatic fluxes 
at 2 keV and 2500 $\AA$ respectively. For the \xmm observation of
2012-12-15 we take the flux at 2500 $\AA$ from the \xmmns-OM UVM2 filter
and find $\alpha_{OX}=-0.96$. This is flatter than usual but still 
consistent, within the 1-$\sigma$ 
error bounds, with the correlation found by \cite{Steffen06} for a sample
of moderate-luminosity optically-selected AGN. Nevertheless, the SED
shows no indication of a 'big-blue-bump' either in the soft X-rays
or in the UV and appears to be relatively X-ray bright. In this respect it 
resembles the SED of a low-luminosity AGN, rather than that of a quasar 
(see \cite{Nemmen}; \cite{ScottStewart}).

%\begin{figure}
%\centering
%\rotatebox{270}{\includegraphics[height=9cm]{sed_0619.ps}}
%\caption[Spectral energy distribution of \srcnamens.]
%{\label{fig:sed} The Galactic-absorption corrected spectral energy distribution of \srcnamens. Data are from ATCA (green), WISE (orange), 2MASS (red),
%\xmm-OM (blue), Galex (pink) and \xmm EPIC-pn (black). The Galex points were
%taken in 2007, likely during a slightly lower luminosity state than the \xmm
%observation.}
%\end{figure}

\section{Discussion}

From the relationship of black hole mass to bulge K-band luminosity
(\cite{MarconiHunt}) we find $M_{BH}\sim10^{7}M_{\odot}$ from the point source
2MASS magnitude of $m_{K}=14.2$. 
From the broad $H_{\alpha}$ luminosity
and width we find a value of 
$M_{BH}\sim3\times10^{7}M_{\odot}$ 
(\cite{greenHo2005}) but note that $L_{H_{\alpha}}$ in \srcname is well below
that of the sources that Greene \& Ho (2005) used in their correlation. 
We can also estimate $M_{BH}$ from the relationship between the broad $H_{\beta}$ line width
and the distance of the BLR gas such that $M_{BH}=fG^{-1}R_{BLR}\Delta\nu^{2}$. From the first AAT spectrum we find $\lambda L_{\lambda} (5100\AA)=10^{43}$ \lumUnits
which, from the correlation and dispersion between $\lambda L_{\lambda}(5100\AA)$ and $R_{BLR}$
for low-luminosity AGN given in \cite{Bentz13} gives $R_{BLR}=10\pm{5}$ light days. From this we obtain $M_{BH}=6\pm{3}\times10^{7}M_{\odot}$.
An $M_{BH}$ of a few $\times10^{7}M_{\odot}$ 
is qualitatively supported by the lack of fast variability
seen in the \xmm light curves. 

At peak, the unabsorbed 0.2-10 keV luminosity was 
$L_{0.2-10}\sim8\times10^{43}$ \lumUnits and 
$L_{2-10}\sim3\times10^{43}$ \lumUnitsns.
This implies a bolometric
luminosity of $L_{bol,peak}\sim7\times10^{44}$ \lumUnits applying the scaling relations
of \cite{Vasudevan09} or Lusso et al. (2011), appropriate for a Sy I. 
The SED implies an unusually
weak big-blue bump in this source and the X-ray bolometric correction seen in 
LLAGN of $\kappa_{X}$$\sim$10 (\cite{Vasudevan07}; \cite{Lusso11}; \cite{Nemmen}) may be more
appropriate here yielding $L_{bol,peak}\sim3\times10^{44}$ \lumUnitsns. 
The ratio of the peak 2--10 keV luminosity to that of [OIII] $\lambda$5007, $L_{X} / L_{OIII} \sim3000$, is in excess of the ratio $\sim 1 - 100$
usually seen in samples of Seyfert galaxies (\cite{Panessa06}; \cite{Lamastra09}).

\subsection{Variability between observations}

Now we investigate the X-ray and UV variability seen between 2012-11-12 and 2013-12-04.
From the X-ray spectra alone it is not simple to distinguish between variable emission and
variable absorption. Here we do not have the benefit of occultation events such 
as those seen in ESO~362-G18 (\cite{Agis}) or ESO~323-G77 (\cite{Miniutti14}) to
help decide the question. If we assume constant emission from the direct
power-law and reflection components in the two XMM-Newton observations
then an increase in absorption of $N_{H}$$\sim$$10^{24}$cm$^{-2}$ and covering
fraction from $\sim$40 to $\sim$70 \% is needed to explain the observed 
flux and spectral change. The extra absorbing cloud(s) must be dusty
to also cause the drop seen in the UV flux and hence must be placed
at or beyond the BLR. However, if the intrinsic emission is also allowed to vary,
then a better fit ($\Delta$C=47 for 1 d.o.f.) is provided by a reduction 
in the power-law emission of a factor 
3 and a {\em reduction} in absorption by 2--10 $\times10^{23}$cm$^{-2}$.
Alternatively, a reduction in intrinsic power-law emission of a factor 2.4 
and a destruction of a disk reflection component also provides a 
better ($\Delta$C=18 for 0 d.o.f.) fit to the joint spectra than a pure increase in absorption
(table~\ref{tab:specfits}). Hence, the spectral evidence favours the idea that
the variability, on a timescale of months, is due to a decrease in intrinsic emission rather than an increase in absorption.

\subsection{Long-term variability}

The factor $\geq$ 140 increase in historical soft X-ray flux
is also difficult to explain by changing line-of-sight absorption.
MKN~335 exhibits large variations in X-ray flux and spectral shape on timescales of
months and years (\cite{Grupe12mkn335}; \cite{Longinotti13}), with 
uncorrelated strong UV flux changes on a similar timescale.
These variations occur in the sense that $L_{X}$ measured at the 
{\it peak} X-ray flux
is consistent with the strong narrow [OIII]$\lambda$5007 luminosity.
Whereas, when the X-ray flux is low it underpredicts $L_{[OIII]}$ and can be said to be
in an X-ray weak state. Similar behaviour is seen in PHL~1092 (\cite{Miniutti09}) although
in that case without parallel UV variability and in PG~2112+059 
(\cite{Schartel}).
Whether the flux responds to varying absorption (\cite{Grupe12mkn335}) or 
a variable disk reflection (\cite{Miniutti09}; \cite{Gallo}) is still an 
open question but in either case the behaviour of 
these AGN is different to that of \srcname, which was X-ray loud when at peak.

Under the assumption then that \srcname has undergone a recent 
increase in luminosity,
we discuss below the likelihood of the change being caused by the 
tidal disruption of a stellar object or by a change in the 
accretion state of a persistent AGN.

\subsection{Tidal-disruption event}

The disruption of a stellar object by gravitational sheer has been invoked to explain
high luminosity flares seen in the nuclei of several quiescent galaxies in the
soft X-ray (\cite{Bade96}; \cite{Komossa99}; \cite{Komossa1242}; \cite{Esquej07}), UV (\cite{Gezari06}; \cite{Gezari09}; \cite{Gezari12}) and optical 
bands (\cite{Komossa09}; \cite{vanVelzen11}; \cite{Cenko}).
In X-ray selected TDE, the X-ray spectrum is usually soft ($\Gamma>3$) and thought to be the Wien tail of a (lightly reprocessed) thermal spectrum which peaks in the EUV. 
Occasionally, tidal disruption candidates do show a harder peak X-ray spectrum more typical of a standard AGN. 
PTF10iya (\cite{Cenko}) showed a strong optical flare accompanied by an X-ray 
flare with a power-law spectrum of slope $\Gamma$=$1.8^{+1.2}_{-1.0}$, similar 
to that seen in \srcnamens. The optical and X-ray spectra of PTF10iya can not be fit simultaneously with a single spectral model, suggesting 
that, while clearly 
part of the same event, if they were produced by a tidal disruption then they represent different reprocessed spectral components of the
original thermal photons. A well-monitored X-ray flare in the nearby 
Sy II galaxy, NGC 4548,
which also resulted in a flat ($\Gamma$$\sim$2.2) power-law spectrum, has also 
been attributed to a tidal disruption event (\cite{Walter13}).
In this case the X-ray flux reduced by a factor $\sim500$ in 18 months
but at peak was a factor $10^{3}$ higher than the value expected from
the relationship between $L_{X}$ and $L_{[OIII]}$. Both of these sources
do show signs of weak AGN activity in their optical spectra.
Transient, super-strong emission lines of Hydrogen, Helium and highly ionized Iron,
 indicated a luminous X-ray flare in \sdssnine (\cite{Komossa08}; \cite{Komossa09}) of
which only the low-energy (UV, optical ,NIR) variability was observed near the high state.
Its X-ray emission, observed a few years after peak was also relatively hard ($\Gamma$$\sim$2).

The flat X-ray spectrum, seen in \srcnamens, initially suggested that we might have seen a similar event to 
\swtd (\cite{Burrows11}; \cite{Bloom}; \cite{Zauderer}), where an on-axis jet dominated the radio and X-ray emission.  
The stringent limits on radio emission, discussed in Sect. 5,
effectively exclude the presence of a strong jet in \srcnamens.
If an {\it off-axis} jet was launched, by the accretion of tidally disrupted
material then it should radiate strongly in the radio band when
it interacts with the interstellar medium and forms a reverse shock
(\cite{Giannios}). The radio emission is
predicted to peak after
$\sim$1 year, at $\nu$$\sim$25 GHz, with a flux of $\sim$$2(D/GPc)^{−2}$ mJy for
a moderately energetic jet . At the distance of \srcname
(330 Mpc) the predicted fluxes at 5.5 and 19 GHz are 10.8 and 16.4 mJy respectively. 
The \srcname 5-$\sigma$ upper
limits taken 14 months after discovery are a factor 98 and 34
below these values which suggests that a jet was not launched
during this event. Note that the strict radio limits also make it unlikely that 
the event was due to a flare in a persistent Blazar.

If \srcname is a TDE then its flux should continue to reduce over the coming years.  

\subsection{High-amplitude variability of AGN}

Variability of AGN by factors of a few to 20 is not uncommon,
and is typically traced back to changes in the line-of-sight 
(cold or ionized) absorption (e.g., \cite{Risaliti05}), 
by changes in emission and reflection mechanisms 
(e.g., \cite{Fabian12}) or by small pertubations in the accretion rate
(\cite{McHardy06}). 

In the X-ray variability selected galaxy, GSN~069, Miniutti et al. (2013) found 
a current accretion rate 25 times higher than the historical value and
a spectrum which appears to be dominated by emission from the accretion
disk suggesting that there has been a change in the accretion disk 
structure within the last 20 years.
Perhaps the closest analogy to the \srcname event is that seen in NGC~2617 (\cite{Shappee}).
This galaxy recently showed an X-ray and nearly-simultaneous UV flare with an
apparent increase in bolometric luminosity of a factor of a few hundred. 
The X-ray spectrum was
flat ($\Gamma=1.7$) and the optical spectrum shows signs of the flare, changing
from a Sy 1.8 classification to that of a Sy 1. 

If \srcname is an AGN, i.e. a SMBH which is being persistently fueled, 
then together with NGC~2617, it represents a relatively fast 
($\lesssim 20$ years) change in the accretion 
rate whose mechanism remains to be explained.

\section{Summary}

A UV and X-ray flare has been seen in the galaxy \srcnamens. The increased luminosity has 
either been caused by a tidal disruption event or 
by an increase
 in the accretion rate of a persistent AGN. The discovery of  
variability events such as \srcnamens, NGC~2617 
and NGC~4548 provide new insights into accretion onto black holes and 
may
eventually lead to an observational framework that can be used to distinguish
between a tidal disruption and AGN variability, in galaxies which do 
show signs of AGN activity in their optical spectra. In this respect, 
surveys like the XSS are important to find further examples of 
these rare events.

\acknowledgements
We thank the XMM OTAC for approving this program.
The XMM-Newton project is an ESA science mission with instruments and contributions directly funded by ESA member states and the USA (NASA).
The \xmm project is supported by the Bundesministerium f\"{u}r Wirtschaft 
und Technologie/Deutches Zentrum f\"{u}r Luft- und Raumfahrt i
(BMWI/DLR, FKZ 50 OX 0001), the Max-Planck Society and the Heidenhain-Stiftung.
We thank the \swift team for approving and performing the monitoring 
observations. This work made use of data supplied by the UK \swift Science Data Centre at the University of Leicester.
%The CSS survey is funded by the National Aeronautics and Space
%Administration under Grant No. NNG05GF22G issued through the Science
%Mission Directorate Near-Earth Objects Observations Program.  
%The CRTS survey is supported by the U.S.~National Science Foundation under
%grants AST-0909182.
We thank the ATCA director for approving our DDT request and the new director of the AAT for granting director's time for the second optical observation. AMR acknowledges the support of STFC/UKSA/ESA funding.
RDS would like to gratefully acknowledge the support and enthusiasm of the late Martin Turner during his introduction to AGN.


\begin{thebibliography}{}
\bibliographystyle{aa}

\bibitem[Agis-Gonzalez et al. 2014]{Agis} Agis-Gonzalez, B., Miniutti, G., Kara, E. et al. 2014, MNRAS (accepted).
\bibitem[Bade, Komossa \& Dahlem 1996]{Bade96} Bade, N., Komossa, S.
\& Dahlem, M. 1996, A\&A, 309, L35 
%\bibitem[Bentz et~al. 2009]{Bentz} Bentz, M., Peterson, B., Pogge, R., \& Vestergaard, M. 2009, ApJ, 694, L166
\bibitem[Bentz et~al. (2013)]{Bentz13} Bentz, M., Denney, K., Grier, C. et al.  2013, ApJ, 767, 149
%\bibitem[Bogdanovic et~al. 2004]{Bogdanovic} Bogdanovic, T., Eracleous, M., Mahadevan, S., Sigurdsson, S., Laguna, P. 2004, ApJ, 610, 707 
%\bibitem[Boller et~al. 1997]{Boller07} Boller, Th.; Brandt, W., Fabian, A. C., Fink, H. 2007, MNRAS, 289, 393 
\bibitem[Bloom et~al. 2011]{Bloom} Bloom, J et al. 2011, Sci, 333, 203
\bibitem[Brandt, Pounds \& Fink 1995]{Brandt} Brandt, W., Pounds, K. \&
 Fink, H., 1995, MNRAS 273, L47.  
%\bibitem[Bruzual \& Charlot 2003]{Bruzual} Bruzual, G. \& Charlot, S. 2003, MNRAS, 344, 1000
\bibitem[Burrows et~al. 2005]{Burrows05} Burrows, D., Hill, J., Nousek, J. et al. 2005, Space Sci. Rev., 120, 165  
\bibitem[Burrows et~al. 2011]{Burrows11} Burrows, D., Kennea, J., Ghisellini, G. et al. 2011, Nat., 476, 421
%\bibitem[Caccianiga et~al. 2007]{cacc07} Caccianiga, A., Severgnini, P., Della Ceca, R. et al. 2007, A\&A, 470, 557
%\bibitem[Cannizzo, Lee \& Goodman 1990]{Cannizzo} Cannizzo, J., Lee, H. \& Goodman, J. 1990, ApJ, 351, 38 
%\bibitem[Cappelluti et~al. 2009]{Cappelluti09} Cappelluti, N., Ajello, M., Rebusco, P. et al. 2009, A\&A, 495, L9
\bibitem[Cash 1979]{Cash} Cash, W. 1979, ApJ 228, 939
\bibitem[Cenko et~al. 2012]{Cenko} Cenko, S., Bloom, J., Kulkarni, S. et al. 2012, MNRAS 420, 2684
%\bibitem[Cenko et~al. 2011]{Cenko11b} Cenko, S., Krimm, H., Horesh, A. et al. 2011, ApJ (arXiv:1107.5307) 
%\bibitem[Chen et~al. 2009]{Chen} Chen, X., Madau, P., Sesana, A., Liu, F., 2009, ApJ, 697, L149 
%\bibitem[Cocchia et~al. (2007)]{Cocchia07} Cocchia, F., Fiore, F., Vignali, C. et al. 2007, A\&A, 466, 31
%\bibitem[Comastri et~al. 2002]{Comastri02} Comastri, A., Mignoli, M., Ciliegi, P. et al. 2002, ApJ, 571, 771
%\bibitem[Done et~al. 2012]{Done12} Done, C., Davis, S., Jin, C., Blaes, O., Ward, M. 2012, MNRAS 420, 1848
%\bibitem[Drake, A.J. et~al. 2009]{catalina} Drake, A.J. et al. 2009, ApJ, 696, 870
%\bibitem[Donley et~al. 2002]{Donley} Donley, J., Brandt, W., Eracleous, M., Boller, Th. 2002, AJ 124, 1308
%\bibitem[Elitzur \& Shlosman 2006]{Elitzur06} Elitzur, M. \& Shlosman, I., 2006, ApJ, 648, L101
\bibitem[Esquej et~al. 2007]{Esquej07} Esquej, P., Saxton, R., Freyberg, M. et al. 2007, A\&A, 462L, 49
\bibitem[Esquej et~al 2008]{Esquej08} Esquej, P., Saxton, R. D., Komossa, S., Read, A. M., Freyberg, M. J., 2008, A\&A, 489, 543 
\bibitem[Evans et~al. (2009)]{Evans} Evans, P., Beardmore, A., Page, K. et al. 2009, MNRAS, 397, 1177 
\bibitem[Fabian et~al. 2002]{Fabian2002} Fabian, A., Ballantyne D., Merloni, A., Vaughan, S., Iwasawa, K. et~al. 2002, MNRAS, 331, L35 
\bibitem[Fabian et~al. 2012]{Fabian12} Fabian, A., Zoghbi, A., Wilkins, D., Dwelly, T., Uttley, P. et al. 2012, MNRAS, 419, 116
%\bibitem[Fossati 2000]{Fossati} Fossati, G., Celotti, A., Chiaberge, M. et al. 2000, ApJ, 541, 166 
\bibitem[Gabriel et~al. 2003]{Gabriel} Gabriel, C. et al. : \emph{}, In ASP Conf. Ser., Vol. 314, ADASS Xiii ed.  Oschenbein, F., Allen, M. \& Egret, D. 759 (2003).
\bibitem[Gallo et al. 2013]{Gallo} Gallo, L., Fabian, A., Grupe, D. et al. 2013, MNRAS, 428, 1191
%\bibitem[Georgantopoulos \& Georgakakis 2005] Georgantopoulos, I. \& Georgakakis, A. 2005, MNRAS, 358, 131
\bibitem[Gierlinski \& Done 2004]{GierlinskDone} Gierlinsky, M. \& Done, C. 2004, MNRAS, 349, L7.
%\bibitem[Gezari et~al. 2003]{Gezari03} Gezari, S., Halpern, J., Komossa, S., Grupe, D., Leighly, K. 2003, ApJ, 592, 42
\bibitem[Gezari et~al. 2006]{Gezari06} Gezari, S., Martin, D., Milliard, B. et al. 2006, ApJ 653L, 25 
%\bibitem[Gezari et~al. 2008]{Gezari08} Gezari, S., Basa, S., Martin, D. et al. 2008, ApJ 676, 944
\bibitem[Gezari et~al. 2009]{Gezari09} Gezari, S., Heckman, T., Cenko, S. et al. 2009, ApJ 698, 1367 
\bibitem[Gezari et~al. 2012]{Gezari12} Gezari, S., Chornock, R., Rest, A et al. 2012, Nat. 485, 217
\bibitem[Giannios \& Metzger 2011]{Giannios} Giannios, D. \& Metzger, B. 2011, MNRAS, 416, 2102
\bibitem[Goad et~al. 2007]{Goad} Goad, M., Tyler, L., Beardmore, A. et al. 2007, A\&A, 476, 1401
\bibitem[Greene \& Ho 2005]{greenHo2005} Greene, J. \& Ho, L. 2005, ApJ, 630, 122
\bibitem[Greiner et~al. 2000]{Greiner} Greiner, J., Schwarz, R., Zharikov, S., Orio, M. 2000, A\&A 362, L25 
\bibitem[Grupe et~al. 1995]{Grupe95} Grupe, D., Beuerman, K., Mannheim, K.
et al. 1995, A\&A 300, L21
\bibitem[Grupe et~al. 1995b]{Grupe95b} Grupe, D., Beuerman, K., 
Mannheim, K. et al. 1995b, A\&A 299, L51
\bibitem[Grupe, Thomas \& Leighly 1999]{Grupe99} Grupe, D., Thomas, H.-C., Leighly, K. 1999, A\&A 350, L31
%\bibitem[Grupe, Komossa \& Gallo 2007]{Grupe07} Grupe, D.,  Komossa, S. \& Gallo, L.2007, ApJ, 668, L111 
%\bibitem[Grupe et~al. 2008]{Grupe08} Grupe, D., Leighly, K.~M., Komossa, S. 2008, AJ, 136, 234
\bibitem[Grupe et~al. 2008]{Grupe08} Grupe, D., Komossa, S., Gallo, L., Fabian, A., Larsson, J. et al. 2008, ApJ, 681,982
\bibitem[Grupe et~al. 2012]{Grupe12mkn335} Grupe, D., Komossa, S., Gallo, L. 2012, ApJS, 199, 28
%\bibitem[Halpern, Gezari \& Komossa 2004]{Halpern04} Halpern, J., Gezari, S. \&
 Komossa, S. 2004, ApJ 604, 572
\bibitem[Hills 1975]{Hills} Hills, J. 1975, Nat. 254, 295
\bibitem[Ho 2008]{Ho08}Ho, L. 2008, Ann. Rev. Astron. Astrophys., 46, 475
\bibitem[Jansen et~al. 2001]{jansen} Jansen, F. et~al, 2001, A\&A, 365, L1
%Giannios, D \& Metzger, B 2011, MNRAS (arXiv:1102.1429) \\
%\bibitem[Jones et~al. 2003]{jones03} 
\bibitem[Just et~al. 2007]{just07} Just, D., Brandt, W., Shemmer, O. et al. 
2007, ApJ, 665, 1004
\bibitem[Kaberla et~al. 2005]{Kaberla} Kalberla, P., Burton, W., Hartmann, D.
 et al. 2005, A\&A, 440, 775
%\bibitem[Karas \& Subr 2007]{KarasSubr07} Karas, V. \& Subr, L. 2007, A\&A, 470, 11 
%\bibitem[Khokhlov \& Melia 1996]{Khokhlov} Khokhlov, A. \& Melia, F. 1996, ApJ 457, L61 
%\bibitem[Komossa 2002]{Komossa02} Komossa, S. 2002, RvMA 15, 27
\bibitem[Komossa \& Bade 1999]{Komossa99} Komossa, S. \& Bade, N. 1999, A\&A, 343, 775 
\bibitem[Komossa \& Greiner 1999]{Komossa99b} Komossa, S. \& Greiner, J. 1999, A\&A, 349, L45 
\bibitem[Komossa et~al. 2004]{Komossa1242} Komossa, S., Halpern, J., Schartel, N. et al. 2004, ApJ, 603, L17
\bibitem[Komossa et~al. 2008]{Komossa08} Komossa, S., Zhou, H., Wang, T. et al.
2008,  ApJ, 678, 13
%\bibitem[Komossa \& Merritt 2008]{KomossaMerritt08} Komossa, S., \& Merritt, D., 2008, ApJ, 683, L21 
\bibitem[Komossa et~al. 2009]{Komossa09} Komossa, S., Zhou, H., Rau, A. et al. 2009, ApJ, 701, 105
\bibitem[Laor 1991]{Laor91} Laor, A. 1991, ApJS, 125, 317
\bibitem[Lamastra et~al. (2009)]{Lamastra09} Lamastra, A., Bianchi, S., Matt, G.
et al. 2009 A\&A, 504, 73
%\bibitem[Lamer, Uttley, McHardy 2003]{Lamer03} Lamer, G., Uttley, P., McHardy, I. M. 2003, MNRAS, 342, L41 
%\bibitem[Lehto \& Valtonen 1996]{LehtoValt} Lehto, H. \& Valtonen, M. 1996, ApJ, 460, L207 
%\bibitem[Leighly et~al. 2009]{Leighly09} Leighly, K., Hamann, F., Casebeer, D. \& Grupe, D. 2009, ApJ, 701, 176 
%\bibitem[Lodato, King \& Pringle 2009]{Lodato09} Lodato, G., King, A.R. \& Pringle, J.E.  2009, MNRAS, 392, 332 
\bibitem[Longinotti et~al. 2013]{Longinotti13} Longinotti, A. L.; Krongold, Y.; Kriss, G. 2013, ApJ, 766, 104
\bibitem[Luminet 1985]{Luminet} Luminet, J.-P. 1985, AnPh, 10, 101
\bibitem[Lusso et~al. 2011]{Lusso11} Lusso, E. et al. 2011, A\&A, 534, 110
%\bibitem[Magdziarz \& Zdziarski 1995]{Magdziarz95} Magdziarz, P. \& Zdziarski, A. 1995, MNRAS, 273, 837
%\bibitem[ Magorrian \& Tremaine 1999]{Magorrian} Magorrian, J. \& Tremaine, S. 
1999, MNRAS 309, 447
%\bibitem[Maksym et~al. 2010]{Maksym} Maksym, W., Ulmer, M., Eracleous, M.,  2010, ApJ 722, 1035
\bibitem[Marconi \& Hunt 2003]{MarconiHunt} Marconi, A., Hunt, L. 2003, ApJ, 589, L21
\bibitem[McHardy et~al. 2006]{McHardy06} McHardy, I., Koerding, E.; Knigge, C.; Uttley, P.; Fender, R. P 2006, Nature, 444, 730 
%\bibitem[Milosavljevic, Merritt \& Ho 2006]{Milosavljevic} Milosavljevic, M., Merritt, D. \& Ho, L. 2006, ApJ 652, 120
\bibitem[Miniutti \& Fabian 2004]{MinFab} Miniutti, G. \& Fabian, A. 2004, MNRAS, 349, 1435
\bibitem[Miniutti et~al. 2009]{Miniutti09} Miniutti, G., Fabian, A., Brandt, W., Gallo, L., Boller, Th., 2009, MNRAS, 396, 85.
\bibitem[Miniutti et~al. 2013]{Miniutti13} Miniutti, G., Saxton, R., Rodrıguez–Pascual, P., Read, A., Esquej, P. et~al. 2013, MNRAS, 433, 1764
\bibitem[Miniutti et~al. 2014]{Miniutti14} Miniutti, G., Sanfrutos, M., Beuchert, T. et~al. 2014, MNRAS, (submitted).
%\bibitem[Nagao, Taniguchi \& Murayama 2000]{Nagao} Nagao, T., Taniguchi, Y. \& Murayama, T. 2000, AJ, 119, 2605
\bibitem[Nandra \& Pounds 1994]{NandraPounds94} Nandra, K. \& Pounds, K. 1994, MNRAS, 267, 193 
\bibitem[Nandra et~al. 2007]{Nandra07} Nandra, K., O'Neill, P., George, I. \& Reeves, J. 2007, MNRAS, 382, 194
%\bibitem[Narayan \& Yi 1994]{Narayan} Narayan, R. \& Yi, I. 1994, ApJ, 428, L13 
\bibitem[Nemmen, Storchi-Bergmann \& Eracleous 2014]{Nemmen} Nemmen, R., Storchi-Bergmann, T. \& Eracleous, M. 2014, MNRAS, accepted.
%\bibitem[Netzer (2009)]{Netzer09} Netzer, H. 2009, MNRAS, 399, 1907
%\bibitem[Nicastro 2000]{Nicastro} Nicastro, F. 2000, ApJ, 530, L65  
%\bibitem[Otani, et~al. 1996]{Otani} Otani, C., Kii, T., Miya, K. 1996, rftu.proc, 491O
\bibitem[Nikolajuk \& Walter 2013]{Walter13} Nikolajuk, M. \& Walter, R. 2013, A\&A, 552, 75
\bibitem[Panessa et~al. 2006]{Panessa06}Panessa, F., Bassani, L., Cappi, M et al. 2006, A\&A, 455, 173
%\bibitem[Phinney 1989]{Phinney} Phinney, E.S., 1989, vol 136 of IAU symposium , 543 
\bibitem[Poole et~al. (2008)]{Poole} Poole, T. et al. 2008, MNRAS, 383, 627
\bibitem[Protassov, van Dyk \& Connors 2002]{Protassov02} Protassov, R., van Dyk, D., \& Connors, A. 2002, ApJ, 571, 545
\bibitem[Read et~al. (2008)]{Read08} Read, A., Saxton, R., Torres, M. et al.
2008, A\&A, 482, L1
\bibitem[Rees 1988]{Rees88} Rees, 1988, Nature 333, 523 
\bibitem[Reeves et~al. 2008]{Reeves08} Reeves, J., Done, C., Pounds, K. et al. 2008, MNRAS, 385, 108
%\bibitem[Rigby et~al. 2006]{Rigby06} Rigby, J., Rieke, G., Donley, J., et al. 2006, ApJ, 645, 115
\bibitem[Risaliti et~al. 2005]{Risaliti05} Risaliti, G., Elvis, M., Fabbiano, G., Baldi, A., Zezas, A. 2005, ApJ, 623, L93 
%\bibitem[Risaliti et~al. 2009]{Risaliti09} Risaliti, G., Miniutti, G., Elvis, M. et al. 2009, ApJ, 696, 160
\bibitem[Roming et~al. 2005]{Roming} Roming, P. et al. 2005, SSRv 120, 95
\bibitem[Ross \& Fabian 2005]{RossFabian} Ross, R. \& Fabian, A. 2005, MNRAS, 358, 211
\bibitem[Sault, Teuben \& Wright 1995]{Sault} Sault, R., Teuben, P. \& 
Wright, M. 1995, ASPC, 77, 433
\bibitem[Saunders et~al. 2004]{Saunders04} Saunders, W., Bridges, T., Gillingham, P., Haynes, R., Smith, G. et al. 2004, SPIE 5492, 389
\bibitem[Saxton et~al. 2008]{Saxton08} Saxton, R., Read, A., Esquej, P. et al.
2008, A\&A 480, 611 
%\bibitem[Saxton et~al. 2011]{Saxton11} Saxton, R., Read, A., Esquej, P., Miniutti, G., Alvarez, E. 2011, In: "Narrow-Line Seyfert 1 Galaxies and Their Place in the Universe", Proceedings of Science, NLS1, 008
\bibitem[Saxton et~al. 2012]{Saxton12} Saxton, R., Read, A., Esquej, P. et al.
2012, A\&A, 541, 106
\bibitem[Schartel et~al. 2010]{Schartel} Schartel, N., Rodríguez-Pascual, P., Santos-Lleo, M. et al. 2010, A\&A, 512, A75
\bibitem[Scott \& Stewart 2014]{ScottStewart} Scott, A.\& Stewart, G. 2014, MNRAS.....
\bibitem[Seyfert 1943]{Seyfert43} Seyfert, C. 1943, ApJ, 97, 28
%\bibitem[Shakura \& Sunyaev 1973]{ShakSun73} Shakura, N. \& Sunyaev, R. 1973, A\&A, 24, 337 
\bibitem[Shappee et~al. 2014]{Shappee} Shappee, B., Prieto, J., Grupe, D.
2014, ApJ, 788, 48
\bibitem[Sharp et al. 2006]{Sharp06} Sharp, R. , Saunders, W., Smith, G., Churilov, V.,  Correll, D. 2006, SPIE, 6296, 14
\bibitem[Sharp \& Birchall 2010]{SharpBirch} Sharp, R. \& Birchall, M. 2010, Publications of the Astronomical Society of Australia, 27, 91
\bibitem[Skrutskie et~al. 2006]{Skrut} Skrutskie M., Cutri, R., Stiening, R., Weinberg, M., Schneider, S. et al. 2006, AJ, 131, 1163
\bibitem[Steffen et~al. 2006]{Steffen06} Steffen, A., Strateva, I., Brandt, W. et al. 2006, AJ, 131, 2826
%\bibitem[Stern \& Laor (2012)]{SternLaor12} Stern, J. \& Laor, A. 2012, MNRAS, 426, 2703
%\bibitem[Strubbe \& Quataert 2009]{Strubbe09} Strubbe, L.E. \& Quataert, E.  2009, MNRAS 400, 2070 
%\bibitem[Strubbe \& Quataert 2011]{Strubbe11} Strubbe, L.E. \& Quataert, E. 2011, MNRAS, 415, 168
\bibitem[Tr\"umper  1983]{Trumper}Tr\"umper J., 1983, Adv. in Space Res. 2, No.4, 241
%\bibitem[Takahashi 1996]{Takahashi} Takahashi, T., et al. 1996, ApJL, 470, L89 
%\bibitem[Trump et~al. 2009]{Trump09} Trump, J. et al. 2009, ApJ, 706, 797 
%\bibitem[Ulmer 1999]{Ulmer99} Ulmer, A. 1999, ApJ, 514, 180  
\bibitem[van Velzen et~al. 2011]{vanVelzen11} van Velzen, S. et al. 2011a, ApJ (arXiv:1009.1627) 
%van Velzen, S. et al. 2011b, MNRAS (arXiv:1104.4105) \\
%\bibitem[Kennedy, Miralda-Escude \& Kollmeier, 2010]{Kennedy10}  \\
\bibitem[Vasudevan \& Fabian 2007]{Vasudevan07} Vasudevan, R.\& Fabian, A. 2007, MNRAS, 381, 1235
\bibitem[Vasudevan et~al. (2009)]{Vasudevan09} Vasudevan, R., Mushotzky, R., Winter, L. \& Fabian, A. 2009, MNRAS, 399, 1553
%\bibitem[Vaughan, Edelson \& Warwick 2004]{Vaughan} Vaughan, S., Edelson, R. \& Warwick, R. 2004, MNRAS, 349, L1 
\bibitem[Veilleux \& Osterbrook 1987]{veillandost} Veilleux, S., \& Osterbrock, D. E. 1987, ApJS, 63, 295
\bibitem[Voges et~al. 1999]{Voges} Voges, W., Aschenbach, B., Boller, T.
et al. 1999, A\&A, 349, 389
%\bibitem[Wang et~al. 2011]{Wang11} Wang, T.-G., Zhou, H.-Y., Wang, L.-F., 
%Lu, H.-L., Xu, D. 2011, ApJ, 740, 85
\bibitem[Wilson et~al. 2011]{Wilson11} Wilson, W., Ferris, R., Axtens, P., Brown, A, David, E. et~al. 2011, MNRAS, 416, 832
\bibitem[Zauderer et~al. 2011]{Zauderer} Zauderer, B. et al. 2011, Nat. 476, 425
%\bibitem[Wang \& Merritt 2004]{WangMerritt} Wang, J. \& Merritt, D. 2004, ApJ, 600, 149
%\bibitem[Yuan et~al. 2010]{Yuan} Yuan, W., Liu, B.~F., Zhou, H. \& Wang, T. 2010, ApJ, 723, 508  
\end{thebibliography}
\end{document}